\documentstyle[prl,aps]{revtex}
\input epsf
\tighten

\begin{document}


\preprint{\vbox{
\hbox{CWRU-P44-2000}
\hbox{DAMTP-2000-47}
}}

\title{Spin and Dualization of SU(5) Dyons}

\author{Tanmay Vachaspati}
\address{
Physics Department,
Case Western Reserve University,
Cleveland OH 44106-7079, U.S.A.
}
\author{Dani\`ele A.\ Steer}
\address{
DAMTP, CMS,
Wilberforce Road, Cambridge CB3 OWA, U.K.
}

\date{\today}

\wideabs{
\maketitle

\begin{abstract}
\widetext
Motivated by the dual standard model, we study the angular
momentum spectrum of stable $SU(5)$ dyons that can be transformed
into purely electric states by a suitable duality rotation {\it
i.e.} are dualizable. The problem reduces to solving a Diophantine
equation for the holomorphic charges in each topological sector,
but the solutions also have to satisfy certain constraints. We
show that these equations can be solved and sets of dualizable,
half-integer spin $SU(5)$ dyons can be found, each of which
corresponds to a single family of the standard model fermions.
We then find two predictions of the dual standard model. 
First, the family of half-integer spin, dualizable dyons
is accompanied by a set of dualizable, integer-spin partner states. 
Secondly, the dyon corresponding to the electron must necessarily
contain non-trivial color internal structure. In addition, 
we provide other general results regarding the spectrum
of dualizable dyons and their novel properties, and extend the
stability analysis of $SU(5)$ monopoles used in the dual standard
model so far to discuss the stability of the half-integer spin
dyons. 
\end{abstract}
}

\narrowtext

\section{Introduction}

The idea that particles may be viewed as solitons
can be traced back to Skyrme \cite{Sky61} who
introduced what is now called the Skyrme model in
which a classical solution (``Skyrmion'') represents
the proton. The model has proved useful in the discussion
of the properties of light nuclei
even though it is known that the Skyrmion does not
have the constituent structure of the proton.

The recent attempts to build a dual standard model
are along the lines that Skyrme developed --- that is, to find
a model that admits soliton analogs of the known
fundamental particles. Partial success in this
direction was achieved in the discovery that the
topological charges of the stable magnetic monopoles
in an $SU(5)$ field theory are in one to one correspondence
with the electric charges of one family of fermions of
the standard model \cite{Vac96,LiuVac97}. A possible scheme
to obtain three families of identically charged magnetic
monopoles was outlined in Ref. \cite{LiuStaVac97}, though at
the expense of considerably complicating the group
structure of the model.

So far, a substantial shortcoming of the model (summarized in
section II) has been that the monopoles emerging from the $SU(5)$
field theory are all bosonic while the standard model particles
are known to be fermionic. The issue of spin and handedness of the
solitons was discussed in Ref. \cite{LiuVac97} though not resolved
in the $SU(5)$ context. The basis for the discussion was the
discovery of ``spin from isospin'' \cite{JacReb76,HasHoo76,Gol76}
in which dyons can have half-integer spin even in a purely bosonic
particle theory. The possibility for handedness was discussed in
the context of a $\theta$ term in the action and the result that
dyons can carry fractional electric charges proportional to
$\theta$ \cite{Wit79}. Also needed was the angular momentum of
dyons in presence of a $\theta$ term \cite{Wil82}.

The success of the spin from isospin phenomenon for the dual
standard model depends on the existence of half-integer
spin states for {\em all} the dyons that ultimately
correspond to the standard model particles. In the case
of the 't Hooft-Polyakov \cite{Hoo74,Pol74} monopole, spin
from isospin can provide half-integer spin to
the fundamental monopole but not to the monopole with
twice the topological winding. In contrast, in the $SU(5)$ case
it has been shown that {\em all} the stable monopoles can be
provided with electric charges to make them into
half-integer spin dyons \cite{Vac98}. However,
the particles that we observe are not ostensibly dyons.
Hence it is important to show that all the half-integer spin
dyons which arise in the $SU(5)$ field theory and which will
be identified with standard model fermions can be transformed
by a duality rotation into purely electric charges. This is
the aim of the present paper.

Here we shall take the approach that the known standard model
particles are purely electric (in contrast to Refs.
\cite{ChaTso98,ChaBorTso99}), and then we would like to know
whether the spin 1/2 dyonic states of the $SU(5)$ model --- that
are in one-one correspondence with the standard model particles
--- can all be dualized into purely electric charges. In trying to
answer this question, we strictly need to consider duality
rotations for gauge fields transforming in representations of the
unbroken symmetry group $H=[SU(3) \times SU(2) \times U(1)]/Z_6$.
Such non-Abelian duality transformations are not fully understood
yet. Our approach (section III) will be to assume that independent
duality rotations can be applied to the field strengths in the
directions of the four commuting generators of $H$. In other
words, we assume that the transformation
\begin{equation}
E^a_i + i B^a_i \rightarrow e^{i\phi_a}
(E^a_i + i B^a_i) \ , \ \ \ a=0,8,3,1
\label{dualityrot}
\end{equation}
is valid with independent phase angles $\phi_a$ for the generators
$\lambda_3$ and $\lambda_8$ of $SU(3)$, $\tau_3$ of $SU(2)$ and
$Y$ of $U(1)$:
\begin{equation}
\lambda_3 = {1\over 2} {\rm diag} (1,-1,0,0,0),
 \label{lambda3}
\end{equation}
\begin{equation}
\lambda_8 = {1\over {2\sqrt{3}}} {\rm diag} (1,1,-2,0,0),
\label{lambda8}
\end{equation}
\begin{equation}
\tau_3 = {1\over 2} {\rm diag} (0,0,0,1,-1),
\label{tau3}
\end{equation}
\begin{equation}
Y = {1\over {2\sqrt{15}}} {\rm diag} (2,2,2,-3,-3).
\label{y}
\end{equation}
The corresponding magnetic charges $m_a$ and electric charges $q_a$
transform as
\begin{equation}
(q_a + i m_a) \rightarrow e^{i\phi_a}(q_a + i m_a).
\label{qm}
\end{equation}
We cannot
rigorously justify these transformations since the $SU(3)$ non-Abelian
equations of motion explicitly involve the gauge fields and not just
the field strengths (in contrast to Maxwell's equations). However,
the equations of motion are indeed invariant under the transformation
if only the commuting gauge field components (Cartan subalgebra)
are non-vanishing. Also note that the Hamiltonian and the Euclidean
action remain invariant under the transformation in (\ref{dualityrot}).
Furthermore, if $SU(5)$ were broken down to $U(1)^4$, each of the
gauge field components labelled by the index $a=0,8,3,1$ would be
Abelian and then the Abelian duality transformations would correspond
exactly to eq.\ (\ref{dualityrot}).

While adopting the transformation in eq.\ (\ref{dualityrot}) as
the duality rotation, we also discuss the case when this may
not be true. If, for example, we restrict ourselves to
$\phi_0 = \phi_8$, we can show that the half-integer spin
states in the even winding topological sectors
must necessarily carry both magnetic and electric $SU(3)$ charge.

In the case when the $\phi_a$ are independent, we find an infinite
set of dualizable, half-integer $SU(5)$ dyon states that are in
one to one correspondence with the standard model particles. 
To arrive at this conclusion we need to solve constrained quadratic
Diophantine equations that can be definite or indefinite. Such
equations have been considered at least since 600 A.D. by Bhaskara
and Brahmagupta and techniques to solve them can be found in
number theory text books ({\it eg.} \cite{Red96}). We shall describe
some of the equations and their solutions in Appendix B.

The infinity of solutions is unlikely to be of any direct
physical relevance. The reason is that we are interested only
in the lowest energy state in any given topological sector since,
presumably, the higher energy states are unstable to decay into
the lowest energy state. However, the energy of a dyon is not
known at strong coupling --- which is the relevant regime for
making contact with the standard model --- and so there is no sure
way of determining the lowest energy states. The best that we
can do at present is to assume a BPS form for the energy
\cite{PraSom75,Bog76}
(see also the monopole reviews in Ref. \cite{bpsreview})
in which the energy of a dyon is proportional to the magnitude
of its charge:
\begin{equation}
E_{BPS} \propto \sqrt{q_a^2 + m_a^2 }
\label{bpsform}
\end{equation}
where a sum over the index $a$ is understood.
This form of the energy does not apply to the dual
standard model where the monopoles may be close to being
BPS but are not exactly BPS \cite{Vac96},
and neither does it apply to the standard model particles.
The purpose of considering eq.\ (\ref{bpsform}) is simply that
it enables us to find the lowest energy dyons in the weak
coupling, near BPS limit.

If we assume eq.\ (\ref{bpsform}) for the energy, then for any given value
of the magnetic charge $m^a$, it would pick out the state $q^a=0$ as
the state with the lowest energy. These purely magnetic states
would have zero spin (see below). The situation is more interesting when
we include a $\theta$ term in the $SU(5)$ action because then the
electric charge contains a contribution from the $\theta$ term
\cite{Wit79}. In that case, the lowest energy state can indeed have
half integer spin. The hope then would be that for a certain value
of $\theta$, of the phases $\phi_a$, and of the coupling constant $g$, one
would obtain a complete family of spin half dyons which would be
the lowest energy states. However, we show that this hope is not
realized due to the monopole with topological winding $n=6$. In
this topological class, the state with the lowest BPS energy necessarily
has integer spin.

Ideally we would like to work with the energy of a dyon
at strong coupling and then determine the lightest states
for given parameters. This would require understanding
the quantum properties of magnetic monopoles --- a subject
that has been under intense research over the last two decades.
Remarkable progress has been achieved in the understanding of
monopoles at strong coupling in the supersymmetric
case \cite{IntSei96} but several tantalizing issues remain open
especially in the non-supersymmetric setting ({\it eg.} \cite{Gol99}).  
An issue that is central to particle-soliton duality is the group
representation in which the monopoles transform when they are
considered as particles. Goddard-Nuyts-Olive conjectured that
monopoles transform in a representation of a dual symmetry
group \cite{GodNuyOli77}. Bais and Schroers \cite{BaiSch98a,BaiSch98b}
find that a richer structure is applicable to non-Abelian monopoles,
since they carry ``holomorphic'' charges in addition to a
topological charge. (This will be important to us in
Sec. \ref{discussion}.) In the $SU(5)$
model, Lepora has provided strong evidence that the monopoles transform
in the fundamental representation of the dual symmetry
group ($SU(3)\times SU(2)\times U(1)$) based on the transformation
properties of the monopoles under rigid gauge transformations
\cite{Lep00a}. This evidence seems to support the concept of
a dual standard model. Further support comes from Lepora's
calculation of the value of the weak mixing angle $\theta_w$
in the context of the $SU(5)$ dual standard model \cite{Lep00b}.
Lepora finds $\sin^2\theta_w =0.22$ which is in good agreement
with experiment at a few GeV. However the relevance of the few
GeV scale to the dual standard model has not yet been investigated.
Naively it seems that this should be the scale at which the
monopole-like structure of elementary particles becomes relevant.
Then it is possible that phenomenological considerations already
impose strong constraints on the idea of the dual standard model.
It would be very interesting to pursue this idea further.

\section{Review of dual standard model}

Consider the symmetry breaking
\begin{equation}
G = SU(5) \longrightarrow
H =  [SU(3)\times SU(2)\times U(1)]/Z_6 \ .
\label{symbreak}
\end{equation}
The magnetic monopoles in this symmetry breaking are labelled by
their $SU(3)$, $SU(2)$ and $U(1)$ magnetic charges,
\begin{equation}
M = (m_0,m_8,m_3,m_1) =
 \biggl ( 0, {{n_8}\over {\sqrt{3}g}},
        {{n_3}\over {2g}}, {- {1}\over {2g}} \sqrt{5\over 3} n_1
                         \biggr )
\label{magcharge}
\end{equation}
where,
\begin{equation}
n_8 = n+3k\ , \ \ \ n_3 = n+2l \ , \ \ \ n_1 =n \ .
\label{nsubas}
\end{equation}
Here, $k$ and $l$ are arbitrary integers since the $\lambda_8$
(of $SU(3)$) and $\tau_3$ (of $SU(2)$) charges are only defined
modulo 3 and 2 respectively.

The topological sector is only determined by the integer $n_1$
which gives the topological winding number ($\Pi_2(G/H) = Z$). The
integers $n_8$ and $n_3$ are related to the ``holomorphic''
charges which are discussed in Ref.
\cite{BaiSch98a,BaiSch98b,Mur89} and which are not topological. In
\cite{Mur89}, Murray derived constraints that the sum of the
topological and holomorphic charges has to be greater than or
equal to zero. The holomorphic charges are the diagonal entries of
the magnetic charge matrix which in this SU(5) case is
\begin{eqnarray}
2 \bf M &=& 2g[m_0\lambda_3 + m_8 \lambda_8 + m_3 \tau_3 + m_1Y ]
                     \nonumber \\
        &=& {\rm diag} ( {{n_8-n_1}\over 3} , {{n_8-n_1}\over 3} ,
                    {{-2n_8-n_1}\over 3}, \nonumber \\
        & & \ \ \ \ \ \ \ \ \ \ \ \ \ \ \
             {{n_3+n_1}\over 2}, {{-n_3+n_1}\over 2} ) \ .
\label{magmatrix}
\end{eqnarray}
Murray's constraints \cite{Mur89} are then that the first three
entries of the charge matrix must be non-negative and the last two
entries must be greater than or equal to minus the topological
charge (our $n_1$).  For $n_1 \le 0$ this leads to:
\begin{equation}
-n_1 \ge 2k \ge 0\ , \ \ \  -n_1 \ge l \ge 0 \ .
\label{klconstraint}
\end{equation}
(For positive values of $n_1$, these inequalities would
be reversed.)

The physical origin of Murray's constraints are, however, not
clear\footnote{In addition, we have not been able to resolve the
different constraints that result if we order the charge matrix
differently. If the charge matrix is organised to have the
$SU(2)$ block first and then the $SU(3)$ block, the constraints in
eq. (\ref{klconstraint}) are reversed and also $2k$ must be
replaced by $k$. Note that Murray's constraints are not
related to the 3-fold color degeneracy and 2-fold spin degeneracy
in the choice of $SU(3)$ and $SU(2)$ generators.}.
As the ingredients in the derivation of the constraints only
involve the structure of the gauge field theory, it is likely that
any configuration that violates the conditions is gauge equivalent
to one that does satisfy the conditions. This is borne out by
considering the trivial case with $n_1 =0$. However, as we shall
see below, the integer $k$ is crucially important in determining
the spin of a dyon:  there are values of $k$ that violate the
constraints but which give rise to angular momentum that cannot be
achieved by states satisfying the constraints. For this reason, we
will assume Murray's constraints provided there is no state that
violates them and which has a different value (integer versus
half-integer) of the angular momentum.

A stability analysis of the non-BPS monopoles in any topological
sector shows that only the $\pm n= 1,2,3,4,6$ monopoles are
stable. (This result assumes a range of parameters in the
$SU(5)$ potential \cite{LiuVac97}.)
A comparison with the standard model particles
shows that these monopoles are in one to one correspondence as
depicted in Table I.

\vspace{0.2cm}

TABLE I.
\small
The quantum numbers ($n_8$, $n_3$ and $n_1$) on stable $SU(5)$
monopoles are shown and these correspond to the $SU(3)$, $SU(2)$
and $U(1)$ charges on the corresponding standard model fermions
shown in the right-most column.
\normalsize
\begin{center}
\begin{tabular*}{8.0cm}{|c@{\extracolsep{\fill}}cccc|}
\hline
&&&&\\[-0.42cm]
                   {$n_{~}$}
                 & {$n_8 /3$}
                 & {$n_3 /2$}
                 & {$n_1 /6$}
                 & {$$} \\
\tableline
&&&&\\[-0.42cm]
+1&1/3&1/2&1/6 &$(u,d)_L$    \\
-2&1/3&0  &-1/3&$d_R$        \\
-3&0  &1/2&-1/2&$(\nu ,e)_L$ \\
+4&1/3&0  &2/3 &$u_R$        \\
-6&0  &0  &-1  &$e_R$        \\
\tableline
\end{tabular*}
\end{center}
\vspace{0.15cm}

The spin of the $SU(5)$ monopoles is provided by bound states
of quanta of a fundamental scalar field. (The existence of such
bound states will depend on the details of the SU(5) potential.
Here we will simply assume that the bound states exist.) This
is the ``spin from isospin'' idea \cite{JacReb76,HasHoo76,Gol76}
extended to $SU(5)$ monopoles \cite{Vac98,LykStr80}. To determine
whether the spin is integer or half-integer, one needs
to calculate the angular momentum in the gauge
fields of the dyon.  It is given by \cite{LykStr80}
\begin{equation}
J = - \sum_a q_a m_a
\label{angmomform}
\end{equation}
where the index $a$ runs over $0, 8, 3, 1$ and labels the two
$SU(3)$ charges, one $SU(2)$ charge and the hypercharge. The $m_a$
have been defined in eq.\ (\ref{magcharge}) and the $q_a$ are the
electric charges present in the state under consideration.

The fundamental scalar field of $SU(5)$ has five components
and we can consider dyonic states with any number of quanta
of these five components. Let us label the components by
the index $h$, then the four different electric charges
on a single quanta of each of the five components can be
written as:
\begin{eqnarray}
e_0^h &= {{g} \over {2}}\pmatrix{1\cr -1\cr 0\cr 0\cr 0\cr} \ , \ \ \
e_8^h = {{g} \over {2\sqrt{3}}}\pmatrix{1\cr 1\cr -2\cr 0\cr 0\cr}\ ,
\nonumber \\
e_3^h &= {{g} \over {2}}\pmatrix{0\cr 0\cr 0\cr 1\cr -1\cr}\ , \ \ \
e_1^h = {{g} \over {2\sqrt{15}}}\pmatrix{2\cr 2\cr 2\cr -3\cr -3\cr}\
.
\label{electriccharges}
\end{eqnarray}
(These assignments are obtained by considering the corresponding
Noether charges.)
To clarify the meaning of these charge assignments consider the
example in which we have one quanta of the first component ($h=1$)
of the fundamental scalar field. This quanta will have $q_0 =g/2$,
$q_8 = g/2\sqrt{3}$, $q_3 = 0$ and $q_1 =g/\sqrt{15}$. Similarly
we can work out the charges on any of the other four ($h=2,3,4,5$)
scalar field components. If we now consider $N_h$ quanta of the
component $h$, then the total electric charge is:
\begin{equation}
Q= (q_0, q_8, q_3, q_1) ,
\label{qnotheta}
\end{equation}
with
\begin{equation}
q_0 = {g\over 2} (N_1-N_2)
\label{q0}
\end{equation}
\begin{equation}
q_8 = {g \over {2\sqrt{3}}} (N_1+N_2-2N_3)
\label{q8}
\end{equation}
\begin{equation}
q_3 = {g \over 2} (N_4-N_5)
\label{q3}
\end{equation}
\begin{equation}
q_1 = {g \over{2\sqrt{15}}} [2(N_1+N_2+N_3)-3(N_4+N_5)].
\label{q1}
\end{equation}

Let us now define
\begin{equation}
M_0 \equiv - (N_1 -N_2)
\label{m0}
\end{equation}
\begin{equation}
M_8 \equiv -(N_1+N_2-2N_3)
\label{m8}
\end{equation}
\begin{equation}
M_3 \equiv -(N_4 - N_5)
\label{m3}
\end{equation}
\begin{equation}
M_1 \equiv -3(N_4+N_5) + 2(N_1+N_2+N_3)) .
\label{m1}
\end{equation}
Since the $N_h$ are integers, so are the $M_a$. Solving
the above equations gives the $N_h$ in terms of the $M_a$:
\begin{equation}
N_2 = N_1 +M_0
\label{n2ma}
\end{equation}
\begin{equation}
N_3 = N_1 + {{M_8+M_0}\over 2}
\label{n3ma}
\end{equation}
\begin{equation}
N_4 = N_1 +{{M_0-M_3}\over 2}+{{M_8-M_1}\over 6}
\label{n4ma}
\end{equation}
\begin{equation}
N_5 = N_1 +{{M_0+M_3}\over 2}+{{M_8-M_1}\over 6}.
\label{n5ma}
\end{equation}
Now since the $N_h$ are integers, we have the following
two constraints on the $M_a$:
\begin{equation}
{{M_8+M_0}\over 2} = {\rm integer}
\label{maconstraint1}
\end{equation}
\begin{equation}
{{M_0-M_3}\over 2}+{{M_8-M_1}\over 6} = {\rm integer} \ .
\label{maconstraint2a}
\end{equation}
The second constraint can be combined with the first to
put it in a more useful form:
\begin{equation}
{{M_8}\over 3}+{{M_3}\over 2}+{{M_1}\over 6} = {\rm integer}\ .
\label{maconstraint2b}
\end{equation}

Then the angular momentum from eq.\ (\ref{angmomform}) with
eqs.\ (\ref{qnotheta}) and (\ref{magcharge}) is found to be:
\begin{equation}
J = {1\over 2} \left [ {{M_8 n_8}\over 3} + {{M_3 n_3}\over 2} +
      {{M_1 n_1} \over 6} \right ].
\label{jmana}
\end{equation}
In Ref. \cite{Vac98} it was shown that we can have
$J=1/2$ for every value of $n$ for suitable values
of the electric charges $M_a$ (which will be different
on the different monopoles). Note that $J$ is only the
angular momentum in the long range gauge fields and does
not contain other possible contributions such as orbital
angular momentum and spin of the gauge particles. These
extra contributions can only change the angular momentum
by an integer and cannot change a half-integer angular
momentum state to one that has integer angular momentum
(or vice versa).

Next let us consider the addition of an $SU(5)$ $\theta$
term. In terms of the gauge fields corresponding to the
diagonal generators, the additional piece of the Lagrangian
is:
\begin{equation}
L_\theta =
\kappa \biggl [ G_{\mu\nu}^3 {\tilde G}^{\mu\nu 3} +
G_{\mu\nu}^8 {\tilde G}^{\mu\nu 8} +
W_{\mu\nu}^3 {\tilde W}^{\mu\nu 3} +
Y_{\mu\nu}   {\tilde Y}^{\mu\nu} \biggr ]
\label{ltheta}
\end{equation}
where,
\begin{equation}
\kappa = {{g^2 \theta}\over {16\pi^2}} \ .
\label{kappa}
\end{equation}
The addition of such a term does not alter the expression for
the angular momentum of the dyons given in eq.\ (\ref{jmana})
but it does affect the values of the electric charges in
eqs. (\ref{q0}-\ref{q1}). (In the case of $SU(2)$ monopoles,
the effect of a $\theta$ term on the electric charge
has been discussed in Ref. \cite{Wit79} and on the angular
momentum of dyons in Ref. \cite{Wil82}.) The new expressions
for the electric charges on the dyons are:
\begin{equation}
q_0 = - {g\over 2}M_0
\label{q0theta}
\end{equation}
\begin{equation}
q_8 = - {g \over {2\sqrt{3}}} M_8 +
           {{g n_8}\over {\sqrt{3}}}{\theta\over{2\pi}}
\label{q8theta}
\end{equation}
\begin{equation}
q_3 = - {g \over 2} M_3 +
           {{g n_3}\over 2}{\theta\over{2\pi}}
\label{q3theta}
\end{equation}
\begin{equation}
q_1 = {g \over{2\sqrt{15}}} M_1 -
           {{g n_1}\over 2}\sqrt{5\over 3}{\theta\over{2\pi}}.
\label{q1theta}
\end{equation}
It is straightforward to check that a shift of $\theta$ by
$2\pi$ can be compensated for by shifts of the $M_a$ that
satisfy eqs. (\ref{maconstraint1}) and (\ref{maconstraint2b}),
thus verifying that the spectrum of states is invariant under
$\theta \rightarrow \theta + 2\pi j$ for any integer $j$.

We now want to know if a complete set
({\it i.e.} all topological sectors occurring in Table 1)
of the half-integer spin dyons can be made to be purely
electric by performing a suitable duality rotation (see Fig. 1).

\begin{figure}
\epsfxsize = 0.9 \hsize  \epsfbox{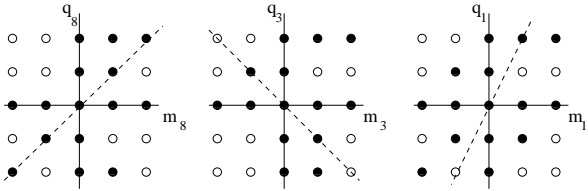}
\vskip 0.5 truecm
\caption{\label{figure1}
The $SU(5)$ dyons have four magnetic and electric charges
and so they can be depicted as points on the four mq-planes
(only three planes have been shown for convenience). We depict
the bosonic states by filled circles and the fermionic states
by unfilled circles. The question
we would like to address is whether a complete family of half-integer
spin dyons lies on straight lines passing through the origin
{\it i.e.} if a rotation of the four sets of axes can give us
a full family of purely electric half-integer spin states.
}
\end{figure}

\section{Dualization of half-integer spin dyons}

The duality rotation phase angles $\phi_a$ (see eq.\ (\ref{dualityrot}))
required to make a dyon into a purely electric object are given by the
inverse tangent of the ratios of its magnetic and electric charges.
Therefore
\begin{equation}
\tan \phi_0 = 0 \ , \ \ \ ({\rm if} \ M_0\ne 0)
\label{tanphi0}
\end{equation}
\begin{equation}
\tan \phi_8 =
- {2\over {g^2}} { 1 \over { M_8/n_8 - 2 \theta /2\pi}}
\label{tanphi8}
\end{equation}
\begin{equation}
\tan \phi_3 =
- {1\over {g^2}} { 1 \over { M_3/n_3 - \theta /2\pi}}
\label{tanphi3}
\end{equation}
\begin{equation}
\tan \phi_1 =
- {5\over {g^2}} { 1 \over { M_1/n_1 - 5\theta /2\pi}}.
\label{tanphi1}
\end{equation}

Note that the $M_a$ are integers and denote the electric charges
on the dyons and hence can depend on the winding number $n$.
Also the integers $n_a$ clearly depend on $n$. For the dyons to
be dualizable, we want that the duality phase angles be independent
of $n$. Hence we require that $\alpha_a$ be independent of $n$
where
\begin{equation}
M_8 = n_8 \alpha_8
\label{m8n8a8}
\end{equation}
\begin{equation}
M_3 = n_3 \alpha_3
\label{m3n3a3}
\end{equation}
\begin{equation}
M_1 =  n_1 \alpha_1
\label{m1n1a1}
\end{equation}

The $\alpha_a$ are independent of $n$ and hence by considering the
dyon with $n_1=1$ we find that $\alpha_1$ must be an integer.
The constraint in eq. (\ref{klconstraint}) shows that we must
also take $n_8 =1$ and $n_3=1$ for $n_1=1$ and so all the
$\alpha_a$ must be taken to be integers\footnote{
Following the discussion after eq.\ (\ref{klconstraint}),
if we relax the constraint to allow $n_8=- 2$ for $n_1=1$,
half-integer values of $\alpha_8$ could still yield integer
values of $M_8$. However, in Appendix A we show that half-integer
values of $\alpha_8$ cannot yield a family of spin half dyons
and so we will restrict our discussion to integer $\alpha_8$.}.
Furthermore, there is a constraint that the $\alpha_a$
must satisfy, coming from the constraint eq.\ (\ref{maconstraint2b})
when combined with eq.\ (\ref{nsubas}) and setting $n=1$:
\begin{equation}
{{\alpha_8}\over 3}+{{\alpha_3}\over 2}+{{\alpha_1}\over 6}
={\rm integer} \ .
\label{alphaconstraint}
\end{equation}

In terms of the $\alpha_a$, the angular momentum is given
by
\begin{eqnarray}
2J_n = \biggl [ {{\alpha_8}\over 3} &+&{{\alpha_3}\over 2} +
               {{\alpha_1}\over 6} \biggr ] n^2 \nonumber \\
   &+&2 [ n ( \alpha_8 k_n + \alpha_3 l_n ) + \alpha_3 l_n^2 ] +
       3 \alpha_8 k_n^2 \ .
\label{2jalpha}
\end{eqnarray}
For the whole family of dyons ($\pm n=1,2,3,4,6$) to have half-integer
spin, we need the right-hand side of eq.\ (\ref{2jalpha}) to be
odd for each member. First consider the $n=2$ monopole. The first term
on the right-hand side is clearly even in this case. The second term
is also even since the $\alpha_a$ are integers.
So $2J_2$ is odd if and only if $3\alpha_8 k_2^2$ is odd. Now
suppose that $\alpha_8$ and $k_2$ are chosen so that
$3\alpha_8 k_2^2$ is odd. Then all the other dyons in the
dualizable family will have half-integer spin if we set $k_n=\pm k_2$
when the first term on the right-hand side of eq.\ (\ref{2jalpha})
is even, and $k_n=0$ when this term is odd.

Two explicit examples satisfying
the constraint in eq.\ (\ref{alphaconstraint}) are:
\begin{equation}
\alpha_8 = 1 \ , \ \ \ \alpha_3 = -1 \ , \ \ \ \alpha_1 = 1 \ .
\label{asoln2}
\end{equation}
\begin{equation}
\alpha_8 = 1 \ , \ \ \ \alpha_3 = 0 \ , \ \ \ \alpha_1 = 4 \ .
\label{asoln3}
\end{equation}
For the first example, the first
term on the right-hand side of eq.\ (\ref{2jalpha}) vanishes and
therefore $2J_n$ is odd provided $k_n^2=$odd for all
$n$\footnote{This is an illustration of the discussion
following eq.\ (\ref{klconstraint}). For $n=1$ the constraint in
eq.\ (\ref{klconstraint}) only allows $k_1=0$. However, $k_1=0$
gives a dyon with integer spin, while the state $k_1=1$ violates
the constraint but gives half-integer spin. Since the spin of the
dyons is not taken into account in deriving the constraints
\cite{Mur89} we assume that the $k_1=1$ state is admissible.}.
Hence a whole family of dyons has half-integer spin and is
dualizable. In fact, there are an infinite number of solutions
($\alpha_a$) that have this property.  This can be seen by noting
that a shift of each of the $\alpha_a$ by any fixed even integer
also leads to a solution that satisfies the constraints and
preserves the half-integer angular momentum.

The dualizable $2J_n=1$ dyon states for a fixed set of $\alpha_a$
correspond to solutions of the Diophantine equation:
\begin{equation}
2\alpha_8 n_8^2 + 3 \alpha_3 n_3^2 = 6 - \alpha_1 n_1^2 \ .
\label{dioph}
\end{equation}
In Appendix B we show that for the $\alpha_a$ in eq.\ (\ref{asoln2})
there are an infinite number of dualizable dyonic states in every
topological sector that have half-integer spin. This conclusion is
expected to be valid whenever some of the $\alpha_a$'s differ in
their signs, leading to indefinite (hyperbolic) Diophantine equations.
If all the $\alpha_a$ have the same sign, we expect there to be
a finite set (possibly empty) of solutions. In view of the
constraints in eq. (\ref{klconstraint}) the infinite set of
states is not of physical interest. Besides we only expect the
lightest of the states for any given winding and angular momentum
to be stable.

\section{General results}
\label{generalresults}

1) {\em There are infinitely many solutions
to the constraints leading to a dualizable family of half-integer
spin dyons}.

This has been shown above in the paragraph following eq.\ (\ref{asoln3}).

\smallskip

2) {\em Each member of the family of dualizable half-integer spin
dyons has an integer spin partner that is also dualizable.}

To see this conclusion, note that if for
a certain $n$ one has $2J_n=$odd, then the state with
$k_n\rightarrow k_n\pm 1$ has (eq. (\ref{2jalpha}))
\begin{equation}
2J_n \rightarrow (2J_n)'=
2J_n + {\rm even\ integer} + 3\alpha_8 \ .
\label{jntojn}
\end{equation}
However, $3\alpha_8$ has to be odd since
$2J_2 = {\rm even}+ 3\alpha_8 k_2^2$ and this has to be
odd for the $n=2$ monopole to have half-integer spin.
Hence the state with the shifted value of $k_n$ has integer
spin.

Hence the dual standard model predicts bosonic partners of
all the standard model fermions. Unlike in the case of
supersymmetry, the masses of the partners do not have to be
degenerate.

\smallskip

3) {\em The $n=6$ dyon with the least BPS energy has integer spin.}

The energy of a BPS dyon is given by eq. (\ref{bpsform}),
\begin{eqnarray*}
E_{BPS} = c \sqrt{q_a^2+m_a^2}
\end{eqnarray*}
where $c$ is a proportionality constant. The state with the
lowest energy is the one with the smallest electric and magnetic
charge. For the $n=6$ dyon, this is the state with $n_8=0=n_3$
since then, both the electric and magnetic charges in the
$SU(3)$ and $SU(2)$ sectors vanish. Now using eq.\ (\ref{jmana})
together with (\ref{m1n1a1}) we see that this state has integer
spin.

A general statement of this kind cannot be made for dyons
with other windings since they necessarily have non-vanishing
$SU(3)$ and/or $SU(2)$ magnetic charge. However it is not difficult
to determine which spin state among the dualizable dyons has the
least BPS energy. First note that dualizability implies
$q_a \propto m_a \propto n_a$. (This relation does not hold for
$a=0$ where we have $q_0 = -gM_0 /2$ and $M_0$ is constrained
by eq. (\ref{maconstraint1}).) Therefore, for fixed values of
the $\alpha_a$, $\theta$ and for small values of $g$ (when the
electric charge contributions are subdominant), the least
BPS energy state is one that has the minimum values of $n_8^2$
and $n_3^2$. For $A\equiv \alpha_8/3+\alpha_3/2+\alpha_1/6=$odd,
this ensures that the $n=1,-2,4$ states with half-integer spin have
lower BPS energy than the corresponding integer spin states.
However, for the $n=-3$ half-integer spin state to have lower
energy than the integer spin state in the case of small $g$,
we need $A=$even because only then the $n_8=0$ ($k_3=1$) state
has half-integer spin.

It is worthwhile pointing out the role of the $\theta$ term in
these considerations. The lowest BPS energy states for a non-zero
$\theta$ angle will occur for non-zero values of the $\alpha_a$.
If $\theta$ were zero, the states with the least energy would be
those with vanishing electric charges (since $\alpha_a =0$ would
minimize the BPS energy) and hence, with zero spin.

\smallskip

4) {\em The $n=2,4,6$ half-integer spin dualizable dyons carry
$\lambda_3$ electric charge {\it i.e.} $M_0 \ne 0$.}

To see this, note that eq.\ (\ref{maconstraint2b}) implies that
$3M_3 + M_1$ is even. Therefore both $M_3$ and $M_1$ are even
or both are odd. For the even $n$ dyons, $M_1 =\alpha_1 n_1$
is even. Hence $M_3$ is also even for even $n$. Now from the
angular momentum formula eq.\ (\ref{jmana}) and the relations
in eq.\ (\ref{nsubas}) we get
\begin{equation}
2J_n = \biggl [ {{M_8}\over 3}+{{M_3}\over 2}+{{M_1}\over 6}
       \biggr ] n + M_8 k_n+ M_3 l_n
\label{jnm}
\end{equation}
Therefore, taking eq.\ (\ref{maconstraint2b}) into account, we see
that $2J_n$ is even for even $n$ if $M_8$ is even.
Hence to obtain an odd value for $2J_n$ ({\it i.e.} half-integer
spin), we must necessarily set $M_8$ to be odd. Next we use the
constraint in eq.\ (\ref{maconstraint1}) which shows that $M_0$
has to be odd and, in particular, has to be non-zero. Therefore
these half-integer states necessarily carry $\lambda_3$ electric
charge.

A consequence of this conclusion is that the two $SU(3)$ duality
rotation phase angles $\phi_0$ and $\phi_8$ cannot be equal.
If non-Abelian duality rotations can only be applied with
$\phi_0=\phi_8$
then the dual standard model would only work provided the particles
transforming non-trivially under $SU(3)$ carry magnetic charge.

\smallskip

5) {\em The $n=6$ half-integer spin dualizable dyon must
have $n_8\ne 0$.}

Inserting $n=6$ in eq.\ (\ref{2jalpha}) shows that we must
necessarily have $k_6=$odd to get half-integer spin.
Therefore $n_8 = n+3k_6 = 3(2+k_6)$ is necessarily non-vanishing
and the $n=6$ half-integer spin state carries SU(3) gluonic
structure.

Similarly if $\alpha_8$ and $\alpha_8/3+\alpha_3/2+\alpha_1/6$ are
odd integers, then $k_3$ has to be even for the $n=3$ monopole
to have half-integer spin. Then $n_8 \ne 0$ and this monopole
also carries gluonic structure.

\section{Stability of half-integer spin dyons}

The monopoles in any topological sector have
two decay channels. Firstly, the monopoles can
emit scalar and vector particles and change
their values of $k$ and $l$. Secondly, a
monopole can fragment into two monopoles of
smaller magnetic charge. We have to show
that neither of these instabilities apply
to the states that we would like to interpret
as standard model particles.

The first instability will not apply to the
lowest lying half-integer spin state in any
given topological sector and so we need only
concern ourselves with the second instability.

Next we show that the dyons with topological winding
$n>6$ are all unstable to fragmentation into dyons
with $n=6$ and something else.

Let us denote the dyonic states by their magnetic
and electric charges as follows:
$$
|n_8,n_3,n_1;M_8,M_3,M_1> \ .
$$
Then we want to show that the decay process:
\begin{eqnarray}
|n_8,n_3&,&n_1; M_8,M_3,M_1> \; \rightarrow \nonumber \\
  &|&n_8,n_3,n_1-6;M_8-p_8,M_3-p_3,M_1-p_1> + \nonumber \\
             ~~~~~~~~~&~&|0,0,6;p_8,p_3,p_1>
\label{decay7}
\end{eqnarray}
is energetically favorable. The two states on
the right-hand side interact by the U(1) magnetic
interactions and we know that this is repulsive.
The electric interactions are small compared to the
magnetic interactions at weak coupling by a factor
$g^4$ and so we ignore them for the present. (Later
we will check that the decay would proceed even with
the electric interactions taken into account.) Hence
it is clear
that this decay process is energetically favorable.
What is not so clear is if the process is allowed by
angular momentum conservation. (The magnetic and
electric charges are conserved in eq.\ (\ref{decay7}).)
This is what we will now check.

The angular momentum of the states on the
right-hand side can be written as (eq.\ (\ref{jmana})):
\begin{equation}
2J_{rhs} = 2J_{lhs} + 2p_3
- M_1
-{{p_8 n_8} \over 3} -{{p_3 n_3} \over 2}
-{{p_1 n_1} \over 6}
\label{jdecay7}
\end{equation}
upto the addition of an integer (which may
be carried off in orbital angular momentum
etc.). For angular momentum conservation
--- meaning that half-integer initial
angular momentum should go to half-integer
final angular momentum and similarly for
integer angular momentum ---  we therefore need
\begin{equation}
- M_1 -{{p_8 n_8} \over 3} -{{p_3 n_3} \over 2}
-{{p_1 n_1} \over 6} = {\rm even}\ {\rm integer}
\label{jconservation7}
\end{equation}
A solution is simply given by $p_8=0=p_3$,
$p_1=6\alpha_1$ because then the left-hand side
is even. With these values of the $p_a$, the electric
interactions are also purely U(1) and repulsive.
This shows that the decay process is not
forbidden by angular momentum conservation and hence
can occur for purely energetic reasons which we know
favor it.

A similar stability analysis goes through for the
$n=5$ dyon.
Consider the decay process:
\begin{eqnarray}
|n_8,n_3&,&5;M_8,M_3,M_1> \; \rightarrow \nonumber \\
          &|&0,n_3,3;p_8,M_3-p_3,p_1> + \nonumber \\
    ~~~~~~&~& |n_8,0,2;M_8-p_8,p_3,M_1-p_1>  .
\label{decay5}
\end{eqnarray}
This is energetically favored since the two dyons
on the right-hand side interact only via the U(1) magnetic
interaction which is repulsive. Next we need to check
if the decay is allowed by angular momentum conservation.

Using the formula for the angular momentum (eq.\ (\ref{jmana})),
we find:
\begin{equation}
2J_{rhs} = 2J_{lhs} -
           \left [ {{n_8 p_8}\over 3} + {{n_3 p_3}\over 2} -
                   {{p_1}\over 6} + {{M_1}\over 2} \right ] .
\label{jdecay5}
\end{equation}
So the decay will be allowed provided:
\begin{equation}
   {{n_8 p_8}\over 3} + {{n_3 p_3}\over 2} -
   {{p_1}\over 6} + {{M_1}\over 2} = {\rm even\ integer} \ .
\label{jconservation5}
\end{equation}
This is clearly so if we choose $p_8=0=p_3$ and
$p_1 =3\alpha_1$. (Recall that $M_1=5\alpha_1$ for the
initial state to be dualizable.) With this choice of
$p_a$, once again the electric interactions are purely
U(1) and repulsive. Hence the $n=5$ dyon is unstable.

This tells us that the dyons with $|n|=5$ and $|n|\ge 7$
are unstable, exactly as found for the monopoles in
\cite{Vac96,GarHar84}. The $\pm n=1,2,3,4,6$ dyons will
still be stable because the fragmentation is completely
governed (in the weak coupling limit) by the magnetic
interactions \cite{Vac96,GarHar84}. Therefore the spectrum
of stable half-integer spin dyons also agrees with the
standard model fermions.

\section{Discussion}
\label{discussion}

Our general results can be found in Sec. \ref{generalresults}.
The main conclusion is that it is possible to find a family
of dyons each member of which has half-integer spin and the
family as a whole can be dualized into purely electric states
(subject to the discussion of duality rotations given in the
introduction). In addition, there are two new features that
have emerged and that may be considered as predictions of
the dual standard model. The first is that each of the
half-integer spin dyons has a bosonic partner. In the dualized
theory, these states would appear as bosonic partners of the
known standard model fermions. Since the bosonic partners are
not due to an imposed symmetry ({\it eg.} supersymmetry),
there is no reason to expect them to be degenerate in mass
with their fermionic partners. The second new feature is that
some of the half-integer spin dyons (in particular the $n=6$
dyon) may have non-vanishing values of $n_8$ and $n_3$ even
though the minimum allowed values of these quantum numbers may
be zero. For example, in the $n=6$ case, the minimum values
are $n_8=0=n_3$, yet to get half-integer spin it is necessary
to have $n_8\ne 0$ (see Sec. \ref{generalresults}). An
interpretation of this result is that since these monopoles with
$n_8 = n+3k_n \ne 0$ carry the same topological charge
as the monopole with winding number $n$ but with $n_8=0$,
they too must transform in the fundamental representation of the
dual symmetry group \cite{Lep00a}. However, the value of $k_n$
is another charge associated with the monopole (related to the
``holomorphic'' charge in \cite{BaiSch98a}).

How is the holomorphic charge manifested in the context of
the dual standard model? 
The holomorphic charge seems to label an internal
degree of freedom of the dualized dyons and, according to 
Bais and Schroers \cite{BaiSch98a}, manifests itself as a 
magnetic dipole moment of the dyons {\it i.e.} an electric dipole 
moment of the particles. Then, for example, the $n=-6$, spin half 
dyon necessarily has $n_8\ne 0$ which means that it must have 
non-trivial $SU(3)$ internal structure even though it transforms 
as an $SU(3)$ singlet. The resolution to this apparent paradox 
is that the particles in the context of the dual standard model 
are composite objects and hence they can have internal $SU(3)$ 
structure in spite of having trivial $SU(3)$ long range interactions 
(as in the case of the proton). The novelty here is that the $n=-6$
dyon under discussion supposedly corresponds to the electron, 
implying that the electron must carry non-trivial internal $SU(3)$
structure!

\

\noindent{\bf Acknowledgements}

We wish to thank Nathan Lepora, Brian Steer,
Mark Trodden and
Serge Winitzki for discussions. We are especially grateful to
Bernd Schroers for help with the constraints derived from
Ref. \cite{Mur89} and to David Singer for number theoretic help.
TV was supported by the DOE.
DAS was supported by PPARC of the UK through a research
fellowship and also the Flora Stone Mather Association
at Case Western Reserve University.


\section*{Appendix A}

Consider the possibility that $\alpha_8$ is a half-integer.
In this case, for $M_8$ to be an integer, $n_8$ should be an even
integer. Then, for even $n$, all of $n_8$, $n_3$ and $n_1$ are
even. Therefore we write $n_a = 2{\tilde n}_a$ where ${\tilde n}_a$
are integers and insert into the equation for the angular momentum
(eq. (\ref{jmana})) to find
\begin{equation}
2J_n = 2 \biggl [ {1\over 3} M_8 {\tilde n}_8 +
                             M_3 {\tilde n}_3 +
                   {1\over 3} M_1 {\tilde n}_1 \biggr ] \ .
\label{jnetc}
\end{equation}
Hence $2J_n$ is even and half-integer spin solutions do not exist.
Therefore half-integer values of $\alpha_8$ cannot yield a family
of half-integer spin dyons.

\section*{Appendix B}

Here we show that there are an infinite number of dyon
states with $J=1/2$ for the choice of $\alpha_a$ in
eq.\ (\ref{asoln2}) (for example).  This is not directly
relevant to us because of eq. (\ref{klconstraint}) and
further physical constraints. However it is still an
interesting exercise.

To see the infinity of solutions, rewrite the angular
momentum constraint (eq.\ (\ref{jmana}) with (\ref{m8n8a8}),
(\ref{m3n3a3}), (\ref{m1n1a1}) and (\ref{asoln2})) as:
\begin{equation}
2n_8^2 - 3n_3^2 = 6-n_1^2 \ .
\label{B1}
\end{equation}
For the fundamental monopole ($n_1 =1$), the problem then is to
find all solutions to the equation
\begin{equation}
2p^2 - 3 q^2 = 5
\label{B2}
\end{equation}
where $p$ and $q$ are integers.

This is a standard problem in number theory and is related to
Pell's equation (for example, \cite{Red96})
The idea of the construction is that given {\it one} solution
to the equation
\begin{equation}
a p^2 - b q^2 = c
\label{B3}
\end{equation}
where $a$, $b$ and $c$ are integers, and if there exists
a non-trivial solution $(l,m)$ to the equation
\begin{equation}
l^2 - ab m^2 =1 \ ,
\label{B4}
\end{equation}
then an infinite set of solutions can be generated. (The trivial
solutions are $l^2=1$, $m=0$.) The construction
uses the solution to the first equation, call it $p_0, q_0$, and the
solution to the second equation, call it $l, m$, to determine another
solution:
\begin{equation}
p= l p_0 + b m q_0 \ , \ \ \ q = a m p_0 + l q_0 \ .
\label{B5}
\end{equation}

So this gives a relatively easy way to check if there are an infinite
number of solutions and to generate them. Indeed for the unit winding
monopole, one can check that this method generates an infinite number
of spin 1/2 states. For the higher winding monopoles, we only need find
one spin 1/2 solution (described below eq.\ (\ref{asoln3})) and that
guarantees an infinite number since
the secondary equation does not care about the value of $c$ and this
is the only place where the topological winding of the monopole ($n_1$)
enters.

In our case we have another restriction on the solutions
$p$ and $q$ since we require $p=n+3k$ and $q=n+2l$ where $k$ and $l$ are
integers. However, it is easy to check that the construction still generates
an infinite sequence of solutions. For the $n=1,2,4$ cases, every alternate
member of the sequence described above has the desired form. For the $n=3,6$
cases, every member has the desired form.

\


\begin{thebibliography}{999}

\bibitem{Sky61} T.H.R. Skyrme, Proc. R. Soc. {\bf A247}, 260 (1958);
{\bf A262}, 237 (1961).

\bibitem{Vac96} T. Vachaspati,
Phys. Rev. Lett. {\bf 76} 188 (1996).

\bibitem{LiuVac97} H. Liu and T. Vachaspati, Phys. Rev. {\bf D56},
1300 (1997).

\bibitem{LiuStaVac97} H. Liu, G.D. Starkman and T. Vachaspati,
Phys. Rev. Lett. {\bf 78}, 1223 (1997).

\bibitem{JacReb76} R. Jackiw and C. Rebbi,
Phys. Rev. Lett. {\bf 36} 1116 (1976).

\bibitem{HasHoo76} P. Hasenfratz and  G. 't Hooft, Phys. Rev. Lett.
{\bf 36} 1119 (1976).

\bibitem{Gol76} A. S. Goldhaber,
Phys. Rev. Lett. {\bf  36} 1122 (1976).

\bibitem{Wit79} E. Witten, Phys. Lett. {\bf B86} 283 (1979).

\bibitem{Wil82} F. Wilczek, Phys. Rev. Lett.  {\bf 48}
1146 (1982).

\bibitem{Hoo74} G. 't Hooft, Nucl. Phys. {\bf B79}
276 (1974).

\bibitem{Pol74} A. M. Polyakov, JETP Lett. {\bf 20} 194 (1974).

\bibitem{Vac98} T. Vachaspati, Phys. Lett. {\bf B427}
323 (1998).


\bibitem{ChaTso98} H. M. Chan and S. T. Tsou,
Phys. Rev. {\bf D57} 2507 (1998).

\bibitem{ChaBorTso99} H. M. Chan, J. Bordes and S. T. Tsou,
Int. J. Mod. Phys. {\bf A14} 2173 (1999).


\bibitem{Red96} ``Number Theory'', by D. Redmound,
Marcel Dekker Inc. (1996).

\bibitem{PraSom75} M.K. Prasad and C. M. Sommerfield,
Phys. Rev. Lett. {\bf 35} 760 (1975).

\bibitem{Bog76} E. B. Bogomolny, Sov. J. Nucl. Phys. {\bf 24}
449-454 (1976).

\bibitem{bpsreview} P. Goddard and  D. I. Olive, Rep. Prog. Phys.
{\bf 41} 1357-1437 (1978);
 S. Coleman, {\it The magnetic monopoles, fifty years later},
in: The Unity of Fundamental Interactions, ed A. Zichichi (Plenum,
London, 1983);
J. Preskill, {\it Vortices and monopoles}, in:
Architecture of Fundamental Interactions at Short Distance, eds P.
Ramond and R. Stora (Elsevier, 1987).

\bibitem{IntSei96} For a review, see K. Intriligator and
N. Seiberg, {\it Lectures on supersymmetric gauge theory
and electric-magnetic duality},
Nucl. Phys. Proc. Suppl. {\bf 45BC} 1 (1996).

\bibitem{Gol99} A. Goldhaber, Phys. Rep. {\bf 315} 83 (1999).

\bibitem{GodNuyOli77} P. Goddard, J. Nuyts, and D. I. Olive,
Nucl. Phys. {\bf B125} 1 (1977).

\bibitem{BaiSch98a} F.A. Bais and B.J. Schroers, Nucl.Phys.
{\bf B512} 250 (1998).

\bibitem{BaiSch98b} F.A. Bais and B.J. Schroers, Nucl.Phys.
{\bf B535} 197 (1998).

\bibitem{Lep00a} N. Lepora, hep-th/0002163.

\bibitem{Lep00b} N. Lepora, hep-ph/0001223.

\bibitem{Mur89} M. Murray, Comm. Math. Phys. {\bf 125}, 661 (1989).

\bibitem{LykStr80} J. D. Lykken and A. Strominger, Phys. Rev. Lett.
{\bf 44} 1175 (1980).

\bibitem{GarHar84} C. Gardner and J. Harvey, Phys. Rev. Lett.
{\bf 52} 879 (1984).


\end{thebibliography}
\end{document}